%% file: main.tex
\documentclass[10pt,twocolumn,letterpaper]{article}

\usepackage[pagenumbers]{cvpr}

\usepackage[T1]{fontenc}
\usepackage[utf8]{inputenc}

\usepackage{amsthm,mathtools}
\usepackage{bm}
\usepackage{dsfont}

\usepackage{multirow}
\usepackage{makecell}
\usepackage{tabularx}
\usepackage{array}
\usepackage{float}
\usepackage{stfloats}   
\usepackage{placeins}   

\usepackage{tikz}
\usetikzlibrary{arrows.meta,positioning,calc,fit,backgrounds}

\usepackage{algorithm}
\usepackage{algpseudocode}

\usepackage{enumitem}
\usepackage{xspace}
\usepackage{siunitx}

\usepackage{url}
\usepackage{xurl}
\usepackage[colorlinks=true,
            linkcolor=blue!55!black,
            citecolor=green!45!black,
            urlcolor=magenta!70!black,
            breaklinks=true,
            bookmarksnumbered=true,
            pdftitle={Training a Compact, Strong, and Realistic Text-to-Speech Model},
            pdfauthor={Audio8-TTS Team},
            pdfsubject={Technical Report},
            pdfkeywords={text-to-speech, discrete audio tokens, dual autoregressive, GRPO,
                         reward model, prosody, realism}
           ]{hyperref}

\def\paperID{0000}
\def\confName{CVPR}
\def\confYear{2026}

\theoremstyle{plain}

\theoremstyle{definition}

\theoremstyle{remark}

\newcommand{\method}{Audio8-TTS\xspace}

\graphicspath{{figures/}{figs/}}

\title{Training a Compact, Strong, and Realistic\\ Text-to-Speech Model}

\author{%
  Runyuan Cai \quad Yu Lin \quad Yiming Wang \quad Xiaodong Zeng \\[0.4em]
  \small \textit{AutoArk}
}
\date{\today}

\begin{document}
\maketitle

\begin{abstract}

Text-to-speech (TTS) systems are conventionally optimized for intelligibility
and speaker similarity. However, neither objective captures the prosodic detail
that distinguishes human performances from synthetic speech, leaving a realism
gap that larger models and higher-quality data can narrow but not close. In the
absence of an objective that directly targets this gap, realism remains a
persistent challenge in TTS. 

We present \method, a compact 0.6B-parameter model for zero-shot voice cloning 
that achieves state-of-the-art performance in intelligibility and speaker similarity, 
with particularly strong results in perceptual naturalness. 
To align the training signal with human perception rather than signal-level fidelity, 
we construct a quality ladder from widely used open-source TTS systems 
and train a realism critic to assess generated speech against matched human recordings 
of the same sentences. 
An identity-erasure objective discourages the critic from relying on generator fingerprints. 
We train the model in three stages: 
multilingual supervised fine-tuning; group relative policy optimization (GRPO) 
with word error rate (WER) and speaker similarity (SIM) rewards; 
and GRPO guided by the realism critic. 
The final stage improves perceived prosodic naturalness and speaker similarity 
while preserving WER performance.
\end{abstract}

\section{Introduction}
\label{sec:intro}

Autoregressive language models over discrete audio tokens have become a leading
paradigm for speech generation. Neural audio codecs encode waveforms as compact
sequences of discrete codes
\citep{zeghidour2021soundstream,defossez2022highfidelity}, enabling language
models trained on these codes to synthesize speech at a quality once attainable
only with dedicated acoustic pipelines \citep{anastassiou2024seedtts}. This
representation also supports unified modeling of recognition, synthesis, and
conversion within a single model \citep{cai2026gpa}, and subsequent work has
extended this line of research to decoupled continuous representations
\citep{shen2026fireredaudio}. 
These advances show that compact speech models can perform demanding audio
tasks when trained with suitable objectives. Accordingly, directly optimizing for
perceptual realism applies this principle to a quality axis absent from
conventional fidelity objectives, enabling compact models to better exploit
their existing capacity. 
How to design a realism oriented training phase remains a key challenge.

TTS systems are commonly evaluated and optimized for two properties: lexical
intelligibility, measured by word error rate (WER), and speaker similarity
(SIM), which measures how closely the synthesized voice matches a reference.
These metrics may appear to cover the main requirements of TTS, but each is a
narrow proxy. WER measures whether the intended words are recognized, not how
naturally they are delivered. SIM measures speaker identity in a compact
representation, leaving much of the utterance's expressive realization
unconstrained. Prosodic details such as delivery style, rhythm, and timing
therefore remain largely outside the score. A model can achieve low WER and
high SIM while still sounding synthetic. Optimizing only these metrics is like
judging athletic ability by bodybuilding standards: it rewards muscularity and
definition, but not necessarily athletic performance. More data and greater
model capacity may narrow this realism gap, but neither directly optimizes
perceptual naturalness. Reinforcement learning, often used to enhance model expressiveness, 
inherits the same limitation when guided only by WER and SIM: it can improve measured
fidelity without necessarily making speech sound more natural.

To address this gap, we present \method, a compact 0.6B-parameter model for
zero-shot voice cloning. Its decoder follows the fast--slow
dual-autoregressive formulation used in recent codec language models
\citep{wang2023valle,liao2024fishspeech}. The slow branch uses Qwen2.5 0.5B
as its backbone, with its vocabulary extended to include audio tokens. A
shallow fast branch predicts the remaining codebooks. 
Together with our training framework, 
this architecture enables \method to achieve state-of-the-art performance in 
intelligibility, speaker similarity,
and perceptual naturalness, despite having far fewer parameters than
substantially larger models.

Training proceeds in three stages: supervised fine-tuning (SFT) on multilingual
speech; group relative policy optimization (GRPO) \citep{shao2024grpo} with WER
and SIM rewards to stabilize content and speaker identity; and a final GRPO
stage guided by extra realism critic to enhance perceptual naturalness. 
We train the critic on a quality ladder of outputs from widely used
open-source TTS systems, ranked through a private blind listening study, with
matched human recordings as the top rung. Unlike public arenas, where each
system speaks in its own voice, our study holds speaker identity and sentence
content constant: every system clones the same reference voice and reads the
same target sentence. The critic scores each TTS candidate against the
matched human recording of that sentence. 
System-specific fingerprints can reveal which generator produced a sample
without revealing how natural it sounds. We therefore introduce an 
identity-erasure strategy to keep the critic's
ratings focused on prosodic realism. 

Another challenge is constraining the output duration within a plausible range. 
Since progressive pause compression is part of our ladder construction strategy,
candidates with less pronounced pauses is designed to receive lower realism scores. 
So optimization can favor longer outputs regardless of the naturalness of speaking rate, 
providing another potential shorcut. Since the WER and SIM are largely insensitive to duration, 
an explicit duration constraint is further introduced as the solution.

In this technical report, we introduce a training framework for compact
zero-shot voice cloning. The framework uses a learned realism critic as a
reward model, complementing WER and SIM rewards with a signal for perceptual
naturalness. Applied to \method, a 0.6B-parameter model, this framework
achieves state-of-the-art results in intelligibility, speaker similarity,
and perceptual naturalness.
\section{Related Work}
\label{sec:background}

\subsection{Neural Text-to-Speech and Discrete Audio Tokens}

Early end-to-end neural text-to-speech systems mapped text to intermediate
acoustic features \citep{wang2017tacotron}; later systems generated waveforms
directly \citep{kim2021vits}. Neural audio codecs introduced another way to
represent speech, encoding waveforms as short sequences of discrete codes
\citep{zeghidour2021soundstream,defossez2022highfidelity}. Self-supervised
speech models also provide discrete units for autoregressive modeling
\citep{hsu2021hubert,lakhotia2021gslm}. Representing speech as a token
sequence makes synthesis a language-modeling task, as in recent zero-shot
systems \citep{anastassiou2024seedtts}. The same approach also allows a single
autoregressive model to handle recognition, synthesis, and conversion through
instruction-driven task induction. We adopt this discrete-token approach and
focus on a compact model for synthesis. Its main design challenges are
efficiently decoding codebook residuals and maintaining perceptual audio
quality.

\subsection{Preference Optimization for Generative Models}

Reinforcement learning from human feedback uses preference signals to align
generative models with human judgments
\citep{christiano2017rlhf,ouyang2022instructgpt}. Proximal policy optimization
(PPO) \citep{schulman2017ppo} is widely used for this purpose, but requires an
auxiliary value model. Group relative policy optimization (GRPO) avoids that
requirement by normalizing rewards across a group of responses to the same
prompt \citep{shao2024grpo}; it is now commonly used to post-train language
models. Applying these methods to speech adds a practical challenge: each
rollout must first be synthesized and then scored. The reward must also
reliably distinguish samples that differ mainly in prosody.

\subsection{Reward Models and Listening Tests for Speech}

Objective quality predictors estimate an absolute mean opinion score (MOS)
from a single utterance \citep{fu2019mosnet}. They can rank degraded samples,
but their training domains are narrow, and they do not directly compare a
candidate with a human performance. Listening arenas compare systems
head-to-head, but each system speaks in its own voice, even when the text is
held fixed. Their rankings can therefore reflect speaker preference as well
as synthesis quality. Our reward model instead scores each candidate against
a matched human recording of the same sentence, holding content and speaker
identity constant. It combines a frozen self-supervised frontend
\citep{chen2022wavlm} with explicit prosodic features, rather than predicting
an absolute quality score.

\section{A Compact Dual-Autoregressive Model}
\label{sec:model}

\begin{figure*}[t]
    \centering
    \includegraphics[width=\textwidth]{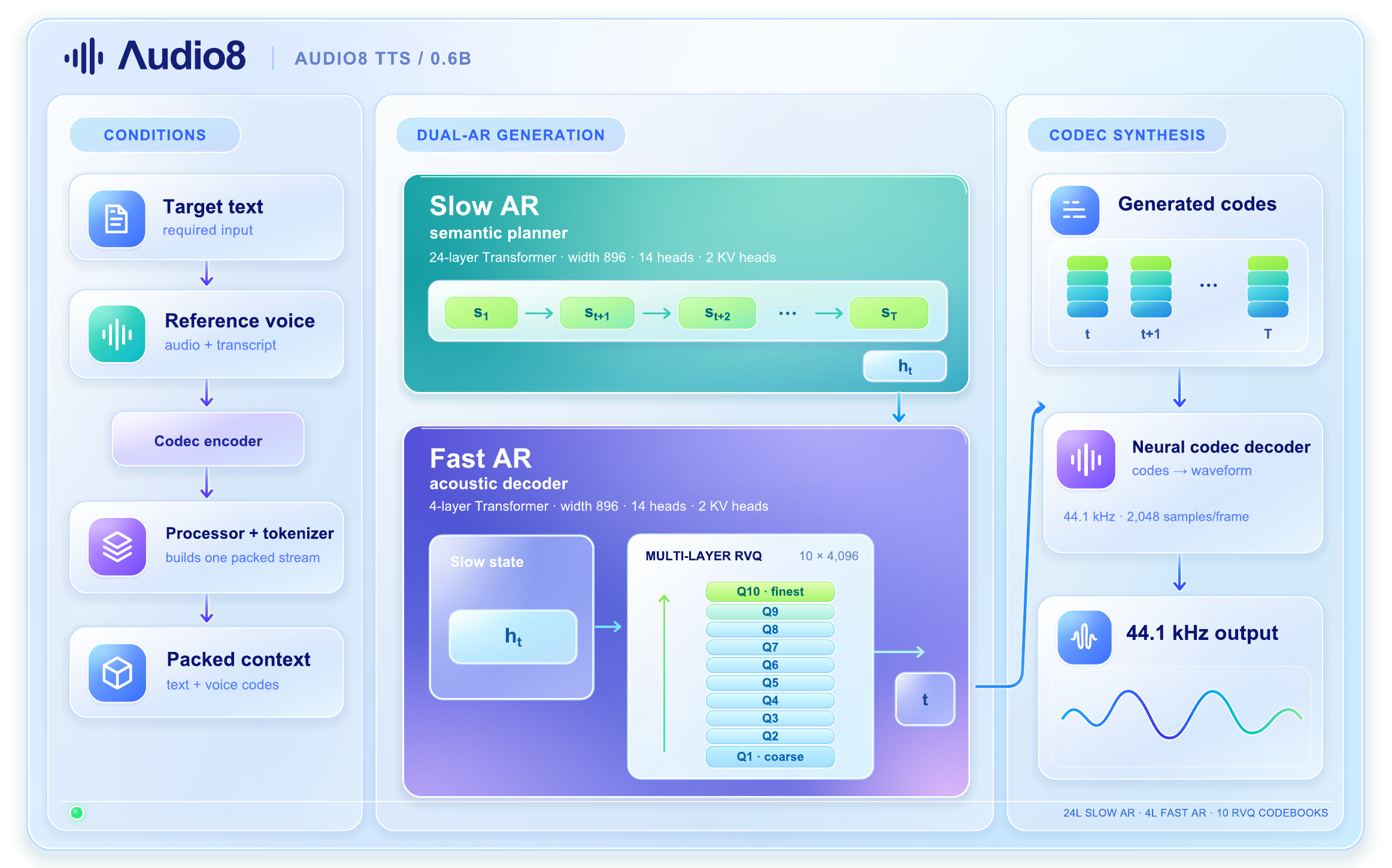}
    \caption{Overview of \method. The slow autoregressive branch predicts text
    tokens and the first audio codebook at the frame rate, while a shallow fast
    branch predicts the remaining nine codebooks in parallel. The audio decoder
    remains frozen.}
    \label{fig:arch}
\end{figure*}

Our decoder adopts the fast--slow dual-autoregressive design used in recent
codec language models \citep{wang2023valle,liao2024fishspeech}. Figure
\Cref{fig:arch} shows the full pipeline. The code, model weights, and implementation and
deployment details are released in the
\href{https://github.com/Edge0-AI/Audio8_TTS}{Audio8-TTS} repository~\citep{audio8tts}. 
We describe the two branches and
their computational cost below.

\subsection{Discrete Audio Tokens}

A fixed audio tokenizer maps each waveform to $T$ frames at
\SI{21.5}{\Hz}. Each frame comprises ten discrete codes, one drawn from each of
ten codebooks with size \num{4096}. Speech synthesis is therefore reduced to
generating a $10 \times T$ matrix of discrete tokens, after which a frozen audio
decoder reconstructs the waveform. Both the tokenizer and the decoder remain
fixed throughout language-model training.

\subsection{Slow and Fast Branches}

Generation is factorized along the frame and codebook axes. The slow branch
operates autoregressively at the frame rate, predicting the text tokens together
with the first audio codebook. It is based on a Qwen2.5 0.5B language model
\citep{qwen2024qwen25} augmented with audio-token embeddings, with a hidden size
of 896 and 24 layers. Conditioned on the hidden state produced by this branch, a
shallow fast branch with four layers predicts the remaining nine codebooks of
each frame in parallel. Because the slow branch operates only at
\SI{21.5}{\Hz}, the autoregressive sequence remains short, while codebook-level
detail is handled by a branch roughly an order of magnitude shallower.

\subsection{Parameter Budget and Inference}

The autoregressive decoder has 601M parameters, excluding the frozen codec.
The slow branch has 24 layers, and the fast branch has 4 layers; both have a
width of 896. They use grouped-query attention with 14 query heads and two
key--value heads. The decoder also includes ten codebook embedding tables,
each with 4096 entries.

At inference, each frame passes through the slow branch once and the fast
branch once. The fast branch predicts nine codebooks in parallel, adding no
autoregressive steps. Thus, sequential computation occurs only at the frame
level.

For CPU inference, the released ONNX implementation uses the same schedule
and applies weight-only INT4 quantization to both autoregressive branches.
Activations, the KV cache, and the codec remain in FP16. In the reported
Apple M2 configuration, memory use is approximately 1.0~GiB after loading
and peaks at 1.1--1.2~GiB. The online model files total about 572~MiB.

\section{Training for Realism}
\label{sec:training}

Training has three stages: multilingual supervised fine-tuning, GRPO with WER
and SIM rewards, and a final GRPO stage guided by the realism critic 
as well as auxiliary WER and SIM reward components.
\subsection{Supervised Fine-tuning}
\label{sec:sft}

For supervised fine-tuning, each example pairs reference and target recordings
from the same speaker and includes a transcript for each. The training set
combines cleaned Emilia \citep{he2024emilia} and YODAS2
\citep{li2024yodas} with a private collection covering speakers and recording
conditions underrepresented in public datasets. We exclude low-quality
recordings, compare transcripts against automatic speech recognition (ASR)
output, and filter samples by language, duration, and ASR error rate. The
target waveform is encoded in the discrete token space defined in
\cref{sec:model}; the reference audio and transcript use the inference-time
conditioning format.

We train both the two branches with cross-entropy: the slow branch predicts text
and the first codebook at the frame rate, while the fast branch predicts the
remaining 9 codebooks in parallel. Training begins with the slow branch
alone, proceeds to the fast branch with the slow branch frozen, and ends with
joint fine-tuning.

This SFT stage yields intelligible, speaker-consistent speech across the
languages in the training corpus. The preference-optimization stages then
preserve and enhance these strengths, while addressing the perceptual unnaturalness that
SFT leaves unresolved (see \cref{sec:intro}).

\subsection{Learning a Reference-Based Realism Critic}
\label{sec:critic}

After SFT, the first GRPO stage optimizes WER and SIM; a second adds a
reference-based reward for perceptual realism. 
We first describe the method for constructing the critic that
provides this reward.

\paragraph{Quality-tier construction.}
A naive human-versus-synthetic classifier cannot rank synthetic speech by
quality. When its synthetic examples come from a single generator, source
identity is confounded with the class label: the classifier can exploit
generator-specific cues rather than learn to assess speech quality.

To provide graded supervision, we construct a quality ladder from outputs
of six open-source systems: Fish-Speech 1.5
\citep{liao2024fishspeech}, IndexTTS-2 \citep{zhou2025indextts2},
Qwen3-TTS \citep{hu2026qwen3tts}, Higgs Audio v3
\citep{bosonai2026higgsv3}, Fish Audio S2
\citep{liao2026fishaudios2}, and Spark-TTS
\citep{wang2025sparktts}. We rank the six TTS systems in a private blind listening study. In each
pairwise comparison, listeners hear outputs for the same sentence and
reference voice, without being told which systems produced them. Holding
content and speaker identity fixed is essential: otherwise, as in public
arenas where systems speak in their own voices, speaker preference can
influence the quality ranking. We aggregate the pairwise judgments into an
Elo-style ranking and place the matched human recording at the top. This
ordered set forms the quality ladder used in our experiments.

All six systems contribute to the listening study and appear on the resulting
quality ladder, but only four provide critic-training data. We reserve Fish
Audio S2\citep{liao2026fishaudios2} and Spark-TTS\citep{wang2025sparktts} for evaluation, 
testing whether the critic can score
unseen systems in a way that preserves their listening-study ranking and
places their scores within the ranges expected for their respective ladder
tiers. 

\paragraph{Critic design.}
The critic scores each synthesized utterance against the matched human
recording of the same sentence. It is trained to rank utterances within
sentence-specific groups, with the human recording anchoring the top of the
quality ladder; its scores do not represent absolute quality.

We pass both the synthesized utterance and its human reference through a
frozen WavLM-large encoder \citep{chen2022wavlm}, extract layer~8, and pool
the frames to 25\,Hz. We also provide explicit pitch, energy, and voicing
trajectories for both. After monotonic-band alignment, multiscale residual
features are mapped to a learned distance $d$. The critic returns the bounded
score
\begin{equation}
s = -B \tanh(d/B), \qquad B = 30,
\end{equation}
where $s \in (-30,0]$ and larger values indicate speech closer to the human
reference. The score is not calibrated as a mean opinion score (MOS). Since
policy optimization normalizes rewards within each group, only relative
scores within a group affect the update.

Only the scoring head is trained. It has roughly 3M parameters; the
316M-parameter WavLM encoder remains frozen. This keeps scoring inexpensive
relative to generating the rollout it evaluates.

We use the same feature-extraction pipeline during training and inference,
including the WavLM layer, temporal pooling, prosodic features, and waveform
preprocessing. Because the critic's scores depend on this representation, we
store a hash of the pipeline configuration with the critic and verify it at
load time. Scoring is disabled if the active configuration differs.

\paragraph{Critic training.}
Each training group is indexed by a sentence and a reference voice. It
contains the matched human recording $y$ and, for each of the $K$ systems
used to train the critic, synthesized utterances
$\{x_{i,\ell}\}_{\ell=0}^{L}$ of the same sentence in the same voice, at
degradation levels $\ell=0,\ldots,L$. The reference voice conditions each
system and fixes speaker identity across systems; the human recording serves
as the perceptual anchor for scoring. For brevity, we write
$s(x_{i,\ell})$ for the reference-conditioned score $s(x_{i,\ell},y)$.

\emph{Constructing the degradation levels.}
We supplement the system-quality ordering with a second comparison axis:
progressively weakened internal rhythm. Here, rhythm refers to the timing and scale 
of auditory pauses. We detect them, scale their durations by a factor, and
resynthesize each utterance using Praat's duration manipulation, based on
pitch-synchronous overlap-add \citep{moulines1990pitch}. 
Other features including pitch, spectral envelope, and
lexical content are preserved during this operation. 
Let $d_j$ denote the original duration of detected pause $j$. At level
$\ell$, its duration is
\[
d_{j,\ell}=a_\ell d_j,
\qquad
1=a_0 \ge a_1 \ge \cdots \ge a_L=0.
\]
Thus, the pause pattern is unchanged at level $0$ and all detected pauses
are removed at level $L$, forming a chain of rhythm degradation. 
To check that the effect is not specific to Praat, we repeat the
construction with WORLD-based resynthesis \citep{morise2016world}. 
Same score trend indicates that the critic's response is not
specific to Praat's resynthesis procedure.

\emph{Monotonicity within a system.}
As pauses are progressively compressed, the score for each system should
decrease by at least a margin $m$:
\begin{equation}
\mathcal{L}_{\mathrm{mono}} =
\sum_{i=1}^{K}\sum_{\ell=0}^{L-1}
\max\!\left(0,\,
s(x_{i,\ell+1}) - s(x_{i,\ell}) + m\right).
\end{equation}
The hinge loss vanishes once the score drop reaches $m$. This margin sets a
minimum score gap between adjacent degradation levels, giving pause-related
changes a local scale that the cross-system ranking loss does not provide.

\emph{Ranking across systems.}
We denote by $\mathcal{P}$ the set of system pairs whose relative quality is
reliably established by the listening study. For each pair in this set, we
train the critic to preserve the observed ordering at every degradation
level:
\begin{equation}
\mathcal{L}_{\mathrm{rank}} =
\sum_{\ell=0}^{L}\sum_{(i,j)\in\mathcal{P}}
\mathrm{BCE}\!\left(
\sigma\!\left(
\frac{s(x_{j,\ell})-s(x_{i,\ell})}{T}
\right),\,
p_{ij,\ell}
\right),
\end{equation}
where $p_{ij,\ell}$ is a soft target derived from the listening-study
quality tiers. As rhythm flattening increases, the target moves toward
$1/2$, allowing the score gap between systems to narrow without reversing
their order. The ranking loss preserves the listening-study order; the
monotonicity loss separately requires scores to fall as each system's rhythm
is flattened.

\emph{Erasure of identity.}
At level $\ell=L$, all detected pauses have been removed. We treat any
remaining score differences across systems as possible generator-identity
leakage and penalize them directly:
\begin{equation}
\mathcal{L}_{\mathrm{era}}=
\sum_{i<j}\left|s(x_{i,L})-s(x_{j,L})\right|.
\end{equation}
The ranking loss may still preserve some score separation at this endpoint,
as specified by $p_{ij,L}$, while $\mathcal{L}_{\mathrm{era}}$ pulls the
scores together. At level $\ell=0$, the ranking target preserves the
listening-study order. In practice, the score spread among fully flattened
variants is a small fraction of that among the original outputs.

The full objective is
\[
\mathcal{L}
=
\mathcal{L}_{\mathrm{rank}}
+\lambda_m\mathcal{L}_{\mathrm{mono}}
+\lambda_e\mathcal{L}_{\mathrm{era}}
+\mathcal{L}_{\mathrm{nuis}},
\]
where $\mathcal{L}_{\mathrm{nuis}}$ penalizes scores predictable from duration
or energy.

\paragraph{Validation.}
We evaluate the critic on 5,600 held-out sentence--voice groups. Each group
contains one utterance from each of the four training systems at each of four
degradation levels, for a total of 89,600 synthesized utterances, in addition
to the matched human references.

\emph{Systems used for training.}
On these held-out groups, the score distributions follow the quality tiers
established by the listening study: the human references occupy the top of
the range, followed by the four systems in ladder order. The critic thus
recovers the listening-study ranking on sentence--voice groups excluded from
training.

\emph{Systems excluded from training.}
Fish Audio S2 and Spark-TTS are not used to train the critic. Their score
distributions nevertheless align with the tiers assigned to them in the
listening study, rather than collapsing into the tiers of the training
systems. The critic therefore preserves the study's quality distinctions
for generators it has not seen during training.

The scoring path is deterministic and performs no resynthesis, so repeated
scoring of the same utterance returns the same value. This consistency is
required during policy optimization, which scores every rollout.

\subsection{Optimizing the Policy with GRPO}
\label{sec:grpo}

The final two training stages use group relative policy optimization (GRPO)
\citep{shao2024grpo}. For each prompt, we sample a group of $N$ rollouts
from the current policy, decode them to audio, and score them. The
stage-specific reward axes are normalized and combined into a scalar $q_i$,
as described below. GRPO then centers and scales this scalar within the
group:
\begin{equation}
A_i=
\frac{q_i-\bar{q}}{\sigma_q+\epsilon},
\qquad
\bar{q}=\frac{1}{N}\sum_{j=1}^{N}q_j,
\qquad
\epsilon=10^{-6},
\end{equation}
where $\sigma_q$ is the standard deviation of the combined scores. The
additive $\epsilon$ stabilizes the normalization when the scores have little
variation.

We then subtract the duration cost to obtain the coefficient used in the
policy update:
\begin{equation}
\tilde{A}_i=A_i-\lambda C_i,
\end{equation}
where $C_i$ is the normalized duration cost defined in
\cref{sec:duration}. The policy update uses $\tilde{A}_i$ to weight the
log-probabilities of rollout $i$. We recompute these log-probabilities with a
forward pass over the sampled sequence rather than reading cached
generation-time values.

The scoring models remain frozen, and no gradient is propagated through them.
During training, scoring runs in a separate, persistent service that transcribes and embeds
decoded audio while the policy samples and updates.

\emph{Fidelity stage.}
The first GRPO stage optimizes lexical accuracy and speaker similarity. We
compute WER with Whisper-large-v3 \citep{radford2023whisper} for non-Chinese
input and an in-house recognizer for Chinese \citep{arkasr2026}. 
We map WER to reward with a clipped linear function, so lower WER yields
higher reward:
\begin{equation}
r_i^{\mathrm{ASR}}
=
\operatorname{clip}\!\left(
1-\frac{\mathrm{WER}_i}{2.0},\,0,\,1
\right),
\end{equation}
where WER is represented as a decimal fraction (e.g., $2\%=0.02$).
We compute SIM as the cosine similarity
between candidate and reference embeddings from a WavLM-based speaker
encoder \citep{chen2022wavlm}, then rescale it to $[0,1]$:
\begin{equation}
r_i^{\mathrm{SIM}}
=
\operatorname{clip}\!\left(
\frac{\mathrm{SIM}_i}{0.85},\,0,\,1
\right).
\end{equation}
The fidelity score combines the two rewards before group-level
standardization:
\begin{equation}
s_i^{\mathrm{fid}}
=
0.7\,r_i^{\mathrm{ASR}}
+
0.3\,r_i^{\mathrm{SIM}}.
\end{equation}

\emph{Realism stage.}
The final GRPO stage combines the fidelity reward with the realism critic's
score. For rollout $i$ in group $g$, let $s_i^{\mathrm{fid}}$ denote its
fidelity score and $s_i^{\mathrm{pros}}$ its critic score. We center each
score by its mean across the group and scale it by the larger of the
group standard deviation and an axis-specific minimum:
\begin{equation}
\begin{aligned}
z_i^{\mathrm{fid}}
&=
\frac{s_i^{\mathrm{fid}}-\bar{s}_g^{\mathrm{fid}}}
{\max(\sigma_g^{\mathrm{fid}},\,0.005)},\\
z_i^{\mathrm{pros}}
&=
\frac{s_i^{\mathrm{pros}}-\bar{s}_g^{\mathrm{pros}}}
{\max(\sigma_g^{\mathrm{pros}},\,0.3)},\\
q_i&=z_i^{\mathrm{fid}}+w\,z_i^{\mathrm{pros}},
\qquad w=1,
\end{aligned}
\end{equation}
where $\bar{s}_g$ and $\sigma_g$ are the mean and standard deviation within
group $g$. The floors prevent small group variances from amplifying either
score. The combined scalar $q_i$ is the reward passed to the GRPO
normalization. With $w=1$, the two axes have equal coefficients, although
the floors can make their effective magnitudes differ.

We combine the axes before computing the GRPO advantage. Computing a
separate advantage for each would normalize each axis independently,
potentially giving a low-variance signal disproportionate influence.
Standardizing the critic score also prevents its prompt-dependent raw spread
from determining its contribution to the combined reward.

\subsection{Bounding Temporal-Scale Drift}
\label{sec:duration}

The pause-compression ladder metioned in \cref{sec:critic} 
provides one controlled cue in critic training:
within each system, progressively compressed pauses receive lower scores.
This teaches the critic to treat pause structure as one aspect of prosodic
realism but can also encourage a favor for longer total duration, 
since preserving pauses increase the length of utterance. 
Hence we constrain temporal scale separately with a one-sided log penalty beyond a
35\% margin over the matched reference. 

Let $\rho_i$ be the ratio of
rollout duration to reference duration, with threshold $k=1.35$. The
one-sided log penalty is
\begin{equation}
u_i=
\operatorname{clip}\!\left(\log\frac{\rho_i}{k},\,0,\,c\right),
\qquad
C_i=
\frac{u_i-\bar{u}_g}{\max(\sigma_g^u,\,0.1)},
\end{equation}
where $c$ caps the penalty, and $\bar{u}_g$ and $\sigma_g^u$ are the mean
and standard deviation of $u_i$ within rollout group $g$. Thus, $u_i=0$
below the duration threshold and increases logarithmically above it. If all
rollouts in a group satisfy the threshold, the centered cost is zero for the
entire group. Dual ascent adjusts $\lambda$ toward a 15\% rate of rollouts
exceeding the threshold. The cost enters the policy update as defined in
\cref{sec:grpo}.

On a fixed evaluation set, duration stays within 6\% of its initial
value, with fewer than 5\% of utterances exceeding the reference by
more than 50\%. Critic scores and speaker similarity nevertheless
improve, supporting gains in delivery rather than duration.

\section{Analysis}
\label{sec:analysis}

We report results on the nine languages in the CV3 benchmark
\citep{du2025cosyvoice}—Chinese, English, Japanese, Korean, German, French,
Spanish, Italian, and Russian—and on thirteen additional languages: Arabic,
Finnish, Hindi, Indonesian, Malay, Dutch, Norwegian, Polish, Portuguese,
Swedish, Thai, Turkish, and Vietnamese. Per-language results are given in
Appendix~\ref{app:perlang}.

\subsection{Data Mixture}
\label{sec:mixture}

Chinese and English each make up about 30\% of the supervised fine-tuning
data; each other language contributes at most 3\%. In the GRPO stage
optimized with WER and speaker-similarity rewards, about two-thirds of the
prompts are in the nine CV3 languages, with roughly 7\% from each. The
remaining prompts are in thirteen additional languages, with roughly 3\%
from each language except Thai, which accounts for about 1\%. For the realism
stage, the prompts corpus uses Chinese and English in roughly equal proportions.

\subsection{Stage-wise Results}

On the nine CV3 languages, mean error rate falls from 5.41\% after SFT to
5.01\% after WER/SIM optimization and 4.86\% after realism optimization.
The WER/SIM stage lowers error in every language. Speaker similarity also
improves across stages, with most of the increase occurring during realism
optimization.

On the thirteen additional languages, WER/SIM optimization yields no
aggregate improvement: error rate rises slightly from 22.36\% to 22.53\%,
while SIM changes from 0.638 to 0.637. 
The aggregate changes are small because improvements in some languages are
offset by regressions in others. By contrast, realism optimization improves both
aggregate metrics, with lower error rates and higher SIM in all thirteen
languages.

\begin{table}[H]
\centering
\footnotesize
\setlength{\tabcolsep}{4pt}
\begin{tabular}{lccc}
\toprule
& SFT & \makecell{GRPO\\(WER/SIM)} & \makecell{GRPO\\(realism)} \\
\midrule
CV3, error rate (\%) $\downarrow$ & 5.41 & 5.01 & \textbf{4.86} \\
CV3, SIM $\uparrow$ & 0.639 & 0.643 & \textbf{0.661} \\
Add. languages$^{\dagger}$, error rate (\%) $\downarrow$ & 22.36 & 22.53 & \textbf{20.77} \\
Add. languages$^{\dagger}$, SIM $\uparrow$ & 0.638 & 0.637 & \textbf{0.679} \\
\bottomrule
\end{tabular}
\caption{Stage-wise results on the nine CV3 languages and thirteen
additional languages. Error rate is WER, except for Chinese, Japanese,
and Korean, where it is CER. The same recognizer and speaker encoder are
used throughout. $^{\dagger}$The additional 13 languages are outside CV3 and
less represented in the training data; their results are reported
separately.}
\label{tab:stages}
\end{table}

\begin{table*}[!t]
\centering
\scriptsize
\setlength{\tabcolsep}{3pt}
\begin{tabular}{lccccccc}
\toprule
& Params.
& \multicolumn{2}{c}{Seed-TTS (en/zh)}
& \multicolumn{2}{c}{CV3 (9)}
& \multicolumn{2}{c}{MiniMax (24)} \\
\cmidrule(lr){3-4}
\cmidrule(lr){5-6}
\cmidrule(lr){7-8}
Model & (B)
& Error $\downarrow$ & SIM $\uparrow$
& Error $\downarrow$ & SIM $\uparrow$
& Error $\downarrow$ & SIM $\uparrow$ \\
\midrule

Seed-TTS \citep{anastassiou2024seedtts}
& $\sim$
& $1.68^{a}$ & $0.779^{a}$
& $\sim$ & $\sim$
& $\sim$ & $\sim$ \\

MiniMax \citep{zhang2025minimaxspeech}
& $\sim$
& $1.24^{a}$ & $0.738^{a}$
& $\sim$ & $\sim$
& $3.77^{a}$ & $0.766^{a}$ \\

IndexTTS-2.5 \citep{li2026indextts25}
& $0.81^{p,\ell}$
& $1.66^{a}$ & $0.852^{a,f}$
& $\sim$ & $\sim$
& $\sim$ & $\sim$ \\

Qwen3-TTS-12Hz-0.6B \citep{hu2026qwen3tts}
& 0.6
& $1.12^{a}$ & $\sim$
& $\sim$ & $\sim$
& $\sim$ & $\sim$ \\

Qwen3-TTS-1.7B \citep{hu2026qwen3tts}
& 1.7
& $1.01^{a}$ & $\sim$
& $7.73^{h}$ & $\sim$
& $27.41^{h}$ & $\sim$ \\

Higgs Audio v3 \citep{bosonai2026higgsv3}
& $4.65^{p}$
& $1.11^{h}$ & $\sim$
& $4.41^{h}$ & $\sim$
& $2.74^{h}$ & $\sim$ \\

Higgs Audio v2 \citep{bosonai2025higgsv2}
& $4.7^{p}$
& $1.17^{e}$ & $0.693^{e}$
& $4.49^{e}$ & $\sim$
& $\sim$ & $\sim$ \\

MOSS-TTS-v1.5 \citep{openmoss2026mosstts15,gong2026mosstts}
& $8.5^{p}$
& $1.52^{c}$ & $0.761^{c}$
& $6.11^{h}$ & $\sim$
& $3.78^{h}$ & $\sim$ \\

FireRedTTS-2 \citep{xie2025fireredtts2}
& 1.7
& $1.72^{h}$ & $\sim$
& $19.20^{h}$ & $\sim$
& $12.52^{h}$ & $\sim$ \\

Fish S2 Pro \citep{liao2026fishaudios2}
& $4.6^{p}$
& $1.32^{e}$ & $0.692^{e}$
& $4.49^{e}$ & $\sim$
& $4.40^{a}$ & $0.780^{a}$ \\

Spark-TTS \citep{wang2025sparktts}
& 0.5
& $1.59^{a}$ & $0.628^{a}$
& $\sim$ & $\sim$
& $\sim$ & $\sim$ \\

CosyVoice3-1.5B \citep{du2025cosyvoice}
& 1.5
& $1.67^{c}$ & $0.750^{c}$
& $6.89^{c}$ & $\sim$
& $\sim$ & $\sim$ \\

VoxCPM2 \citep{zhou2026voxcpm2}
& $2.3^{p}$
& $1.41^{c}$ & $0.774^{c}$
& $5.35^{c}$ & $\sim$
& $\sim$ & $\sim$ \\

\midrule
\method{} (ours)
& 0.6
& $1.25^{a}$ & $0.704^{a}$
& 4.86 & 0.661
& 5.89 & 0.754 \\
\bottomrule
\end{tabular}

\caption{Results on Seed-TTS, CV3, and MiniMax.
Error rates (\%) are macro-averaged over languages.
Seed-TTS averages English WER and Chinese CER, excluding zh-hard.
$\sim$: unavailable.
$^{a}$Means computed from reported per-language results.
$^{c}$Results compiled in the VoxCPM repository
\citep{openbmb2026voxcpm}.
$^{e}$Our evaluation.
$^{h}$Boson AI's evaluation \citep{bosonai2026higgsv3}.
$^{f}$FunASR-based SIM.
$^{p}$Parameter counts computed from checkpoint tensors,
excluding codec weights and buffers; otherwise, counts are taken
from model releases.
$^{\ell}$Excludes the separate acoustic decoder.
Parameter-counting conventions differ across models.}
\label{tab:suites}
\label{tab:suites_multilingual}
\end{table*}

In addition to the nine-language CV3 benchmark, we evaluate the final
model on Seed-TTS \citep{anastassiou2024seedtts} and the 24-language
MiniMax test set \citep{zhang2025minimaxspeech}.
\Cref{tab:suites} compares our results with other TTS systems.

We reevaluate Fish S2 Pro on Seed-TTS and CV3 because its reported
results use a different text-normalization procedure. We also
evaluate Higgs Audio v2, for which the required results were
not available.
For the remaining baseline results, we use the original reports,
Boson AI's evaluation \citep{bosonai2026higgsv3}, and results
compiled in the VoxCPM repository \citep{openbmb2026voxcpm}.
Evaluation protocols differ across sources; in particular, SIM
scores computed with different speaker encoders are not directly
comparable. Full-suite results may include languages outside a
model's supported set.

For Qwen3-TTS, the Seed-TTS results use the 12Hz Base models.
Boson's CV3 and MiniMax results do not specify the tokenizer variant.
Per-language CV3 error rates for these systems, including the two hard
subsets, are reported in \Cref{tab:cv3_detail}.

Across the 22 CV3 languages, the final model has higher speaker similarity
than the SFT model in every language. Recognition error is lower in all but
Korean and Thai; for both, it remains within one percentage point of the
post-SFT result (\Cref{tab:cv3perlang}). 
On MiniMax, Cantonese is the only language with an error rate above
20\% (\Cref{tab:mmperlang}). This likely reflects the training data's
strong emphasis on Mandarin Chinese, as well as the substantial overlap
between the token spaces for Cantonese and Mandarin.

\subsection{Blind Human Evaluation}

We conducted a blind listening study to compare outputs from the SFT,
WER/SIM, and realism stages. The realism-stage model received a
substantially higher Elo rating than either of the other two models,
while the SFT and WER/SIM models received similar ratings.

\section{Limitations and Conclusion}
\label{sec:conclusion}

\emph{Limitations.}
The realism critic requires a matched human reference for each sentence. Its
scores are relative: they rank candidates for the same prompt according to
the ordering established by the listening study, but do not provide an
absolute measure of naturalness or replace a content check. 
Though the critic ranks overall synthesis quality across systems, its
controlled signal for separating prosodic cues from generator identity
comes from pause-flattened variants. These variants isolate pause timing;
they do not provide equally targeted supervision for pitch, energy, or
other prosodic cues.

The duration penalty reduces length drift but does not eliminate it.
Preference optimization has a different limit: it can make existing
prosodic patterns more or less likely, but cannot teach the model patterns
it never produces.

The 13 additional languages account for a smaller share of the
training data, and their recognition error remains substantially higher
than that on the nine CV3 languages. We do not evaluate audio artifacts
such as buzzing or vocoder noise and the metrics used 
in evaluation would not capture them.

\emph{Conclusion.}
We presented Audio8-TTS, a 0.6B dual-autoregressive model trained in three
stages: supervised fine-tuning, GRPO with word error rate and speaker
similarity rewards, and GRPO with a reference-anchored realism critic. The
critic is trained against a quality ordering established through a blind
listening study, with an identity-erasure objective to keep its score from
tracking generator identity.

On the nine CV3 languages, error rate falls at each stage and mean speaker
similarity rises. On the 13 additional languages, the WER/SIM stage
leaves aggregate results nearly unchanged, while the realism stage improves
both metrics in every language, despite using only Chinese and English
prompts. In a separate blind listening study, the realism-stage model
receives a substantially higher Elo rating than the SFT and WER/SIM models,
whose ratings are similar. The duration constraint also keeps output
duration close to its initial level as critic scores and speaker similarity
improve.

\section*{Acknowledgments}

We thank the developers of Fish-Speech 1.5 \citep{liao2024fishspeech},
IndexTTS-2 \citep{zhou2025indextts2}, Qwen3-TTS \citep{hu2026qwen3tts},
Higgs Audio v3 \citep{bosonai2026higgsv3}, Fish Audio S2
\citep{liao2026fishaudios2}, and Spark-TTS \citep{wang2025sparktts} for
releasing the systems used in our quality ladder. The Fish
Audio S2 implementations also underpin much of our training and inference
code.

We thank the CosyVoice 3 \citep{du2025cosyvoice} and Seed-TTS
\citep{anastassiou2024seedtts} teams for their evaluation suites, and the
authors of Whisper \citep{radford2023whisper} and WavLM
\citep{chen2022wavlm} for releasing the models used for scoring.
\FloatBarrier   
\bibliographystyle{ieeenat_fullname}
\bibliography{refs}

\clearpage
\onecolumn          
\appendix
\section{Per-Language Results}
\label{app:perlang}

\begin{center}
\scriptsize
\renewcommand{\arraystretch}{0.92}
\setlength{\tabcolsep}{4pt}
\input{tab_cv3_perlang_2col.tex}
\captionof{table}{CV3 per-language results on the nine benchmark languages and
the thirteen additional languages, for the model after supervision and after
training. Error rates are word error rates, except for Chinese, Japanese and
Korean, where they are character error rates. $^{\dagger}$ marks the languages we
added, which lie outside the benchmark's nine.}
\label{tab:cv3perlang}
\end{center}

\begin{center}
\scriptsize
\renewcommand{\arraystretch}{0.92}
\setlength{\tabcolsep}{6pt}
\input{tab_minimax_perlang_2col.tex}
\captionof{table}{MiniMax per-language results on the 24-language multilingual
set, for the final model. Error rates are word error rates, and character error
rates for the languages the suite scores by character: Chinese, Cantonese,
Japanese, Korean, Arabic, Greek, Hindi, Thai and Vietnamese.}
\label{tab:mmperlang}
\end{center}
\begin{center}
\scriptsize
\renewcommand{\arraystretch}{0.92}
\setlength{\tabcolsep}{4pt}
\begin{tabular}{lcccccc}
\toprule
& \multicolumn{2}{c}{Seed-TTS, en}
& \multicolumn{2}{c}{Seed-TTS, zh}
& \multicolumn{2}{c}{Seed-TTS, zh-hard} \\
\cmidrule(lr){2-3}
\cmidrule(lr){4-5}
\cmidrule(lr){6-7}
Model
& WER $\downarrow$ & SIM $\uparrow$
& CER $\downarrow$ & SIM $\uparrow$
& CER $\downarrow$ & SIM $\uparrow$ \\
\midrule

Seed-TTS \citep{anastassiou2024seedtts}
& 2.25 & 0.762
& 1.12 & 0.796
& 7.59 & 0.776 \\

MiniMax \citep{zhang2025minimaxspeech}
& 1.65 & 0.692
& 0.83 & 0.783
& $\sim$ & $\sim$ \\

IndexTTS-2.5 \citep{li2026indextts25}
& 1.89 & $0.855^{f}$
& 1.43 & $0.848^{f}$
& $\sim$ & $\sim$ \\

Qwen3-TTS-12Hz-0.6B \citep{hu2026qwen3tts}
& 1.32 & $\sim$
& 0.92 & $\sim$
& $\sim$ & $\sim$ \\

Qwen3-TTS-12Hz-1.7B \citep{hu2026qwen3tts}
& 1.24 & $\sim$
& 0.77 & $\sim$
& $\sim$ & $\sim$ \\

Fish S2 Pro \citep{liao2026fishaudios2}
& $1.607^{e}$ & $0.646^{e}$
& $1.038^{e}$ & $0.738^{e}$
& $10.149^{e}$ & $0.701^{e}$ \\

Higgs Audio v2 \citep{bosonai2025higgsv2}
& $1.524^{e}$ & $0.664^{e}$
& $0.806^{e}$ & $0.721^{e}$
& $10.622^{e}$ & $0.693^{e}$ \\

MOSS-TTS-v1.5 \citep{openmoss2026mosstts15,gong2026mosstts}
& $1.85^{c}$ & $0.734^{c}$
& $1.20^{c}$ & $0.788^{c}$
& $\sim$ & $\sim$ \\

CosyVoice3-1.5B \citep{du2025cosyvoice}
& $2.22^{c}$ & $0.720^{c}$
& $1.12^{c}$ & $0.781^{c}$
& $5.83^{c}$ & $0.758^{c}$ \\

VoxCPM2 \citep{zhou2026voxcpm2}
& $1.84^{c}$ & $0.753^{c}$
& $0.97^{c}$ & $0.795^{c}$
& $8.13^{c}$ & $0.753^{c}$ \\

Spark-TTS \citep{wang2025sparktts}
& 1.98 & 0.584
& 1.20 & 0.672
& $\sim$ & $\sim$ \\

\midrule
\method{} (ours)
& 1.46 & 0.663
& 1.03 & 0.745
& 12.80 & 0.702 \\
\bottomrule
\end{tabular}

\captionof{table}{Seed-TTS results by split.
WER and CER are reported as percentages.
The zh-hard split is excluded from the means in \Cref{tab:suites}.
$^{c}$Results compiled in the VoxCPM repository
\citep{openbmb2026voxcpm}.
$^{e}$Our evaluation.
$^{f}$FunASR-based SIM.
SIM scores computed with different speaker encoders are not
directly comparable.
$\sim$: unavailable.
Only systems with reported per-split results are included.
The Seed-TTS model's zh-hard results are taken from
\citet{du2025cosyvoice}.}
\label{tab:suites_seed_detail}
\end{center}

\begin{center}
\scriptsize
\renewcommand{\arraystretch}{0.92}
\setlength{\tabcolsep}{4pt}
\begin{tabular}{lcccccccccccc}
\toprule
Model & Params (B)
& zh & en & hard-zh & hard-en & ja & ko & de & es & fr & it & ru \\
\midrule

Fish S2 Pro \citep{liao2026fishaudios2}
& 4.6
& $3.600^{e}$ & $3.493^{e}$ & $10.588^{e}$ & $7.349^{e}$ & $5.139^{e}$
& $4.111^{e}$ & $3.605^{e}$ & $2.972^{e}$ & $8.600^{e}$ & $4.229^{e}$ & $4.702^{e}$ \\

Higgs Audio v2 \citep{bosonai2025higgsv2}
& 4.7
& $3.378^{c}$ & $3.404^{c}$ & $10.424^{c}$ & $5.754^{c}$ & $4.742^{c}$
& $4.260^{c}$ & $3.300^{c}$ & $2.929^{c}$ & $9.425^{c}$ & $3.555^{c}$ & $5.423^{c}$ \\

CosyVoice3-1.5B \citep{du2025cosyvoice}
& 1.5
& $3.91^{c}$ & $4.99^{c}$ & $9.77^{c}$ & $10.55^{c}$ & $7.57^{c}$
& $5.69^{c}$ & $6.43^{c}$ & $4.47^{c}$ & $11.8^{c}$ & $10.5^{c}$ & $6.64^{c}$ \\

VoxCPM2 \citep{zhou2026voxcpm2}
& 2.3
& $3.65^{c}$ & $5.00^{c}$ & $8.55^{c}$ & $8.48^{c}$ & $5.96^{c}$
& $5.69^{c}$ & $4.77^{c}$ & $3.80^{c}$ & $9.85^{c}$ & $4.25^{c}$ & $5.21^{c}$ \\

\midrule
\method{} (ours)
& 0.6
& 1.814 & 2.907 & 10.665 & 4.736 & 7.511
& 4.801 & 3.200 & 3.389 & 7.938 & 4.933 & 7.258 \\
\bottomrule
\end{tabular}

\captionof{table}{Per-language CV3 error rates (\%): word error rates, except
for Chinese, Japanese, Korean and the Chinese hard subset, which are character
error rates. $^{c}$Results compiled in the VoxCPM repository
\citep{openbmb2026voxcpm}. $^{e}$Our evaluation. Each language contributes 500
items, except the two hard subsets (60 Chinese, 64 English). The mean over the
nine benchmark languages is the CV3 (9) column in \Cref{tab:suites}.}
\label{tab:cv3_detail}
\end{center}

\end{document}

%% file: tab_cv3_perlang_2col.tex
\begin{tabular}{lcccc@{\hspace{14pt}}lcccc}
\toprule
& \multicolumn{2}{c}{Error rate $\downarrow$} & \multicolumn{2}{c}{Similarity $\uparrow$} & \multicolumn{2}{c}{Error rate $\downarrow$} & \multicolumn{2}{c}{Similarity $\uparrow$} \\
\cmidrule(lr){2-3}\cmidrule(lr){4-5}\cmidrule(lr){7-8}\cmidrule(lr){9-10}
Language & SFT & final & SFT & final & Language & SFT & final & SFT & final \\
\midrule
Chinese & 2.03 & 1.81 & 0.668 & 0.690 & Hindi$^{\dagger}$ & 22.91 & 21.71 & 0.685 & 0.725 \\
English & 3.03 & 2.91 & 0.610 & 0.629 & Indonesian$^{\dagger}$ & 15.56 & 13.74 & 0.629 & 0.668 \\
Japanese & 8.10 & 7.51 & 0.649 & 0.666 & Malay$^{\dagger}$ & 19.48 & 17.80 & 0.636 & 0.671 \\
Korean & 4.60 & 4.80 & 0.681 & 0.696 & Dutch$^{\dagger}$ & 11.78 & 10.87 & 0.652 & 0.692 \\
German & 3.64 & 3.20 & 0.646 & 0.667 & Norwegian$^{\dagger}$ & 20.61 & 17.96 & 0.635 & 0.691 \\
French & 8.66 & 7.94 & 0.616 & 0.635 & Polish$^{\dagger}$ & 21.68 & 19.23 & 0.635 & 0.674 \\
Spanish & 4.89 & 3.39 & 0.643 & 0.668 & Portuguese$^{\dagger}$ & 7.89 & 6.87 & 0.640 & 0.677 \\
Italian & 5.54 & 4.93 & 0.619 & 0.650 & Swedish$^{\dagger}$ & 20.97 & 19.25 & 0.671 & 0.710 \\
Russian & 8.21 & 7.26 & 0.621 & 0.645 & Thai$^{\dagger}$ & 47.97 & 48.57 & 0.626 & 0.669 \\
Arabic$^{\dagger}$ & 23.17 & 21.10 & 0.571 & 0.612 & Turkish$^{\dagger}$ & 13.38 & 12.57 & 0.612 & 0.650 \\
Finnish$^{\dagger}$ & 33.80 & 32.28 & 0.663 & 0.703 & Vietnamese$^{\dagger}$ & 31.48 & 28.11 & 0.638 & 0.679 \\
\bottomrule
\end{tabular}

%% file: tab_minimax_perlang_2col.tex
\begin{tabular}{lcc@{\hspace{22pt}}lcc}
\toprule
Language & Error rate $\downarrow$ & Similarity $\uparrow$ & Language & Error rate $\downarrow$ & Similarity $\uparrow$ \\
\midrule
Arabic & 2.40 & 0.704 & Italian & 1.96 & 0.737 \\
Cantonese & 31.35 & 0.771 & Japanese & 5.14 & 0.789 \\
Chinese & 1.11 & 0.795 & Korean & 3.73 & 0.785 \\
Czech & 8.63 & 0.733 & Polish & 1.96 & 0.789 \\
Dutch & 1.72 & 0.705 & Portuguese & 2.23 & 0.786 \\
English & 2.16 & 0.774 & Romanian & 17.43 & 0.733 \\
Finnish & 5.07 & 0.791 & Russian & 3.07 & 0.749 \\
French & 3.20 & 0.666 & Spanish & 1.35 & 0.745 \\
German & 0.90 & 0.751 & Thai & 6.67 & 0.776 \\
Greek & 4.18 & 0.791 & Turkish & 1.56 & 0.765 \\
Hindi & 9.21 & 0.817 & Ukrainian & 16.34 & 0.694 \\
Indonesian & 2.13 & 0.727 & Vietnamese & 7.97 & 0.712 \\
\bottomrule
\end{tabular}